\documentclass[twocolumn]{aastex62}

\usepackage{textcomp}
\usepackage{amsmath}
\usepackage{float} 
\usepackage{subfig}

\received{2 Sep 2020}
\accepted{4 Jan 2021}

\submitjournal{ICARUS}

\begin{document}

\title{\Large Experimental study of clusters in dense granular gas and implications for {the particle stopping time} in protoplanetary disks}

\correspondingauthor{Niclas Schneider}
\email{niclas.schneider@uni-due.de}

\author{Niclas Schneider}
\affil{University of Duisburg-Essen, Faculty of Physics, Lotharstr. 1-21, 47057 Duisburg, Germany}

\author{Grzegorz Musiolik}
\affil{University of Duisburg-Essen, Faculty of Physics, Lotharstr. 1-21, 47057 Duisburg, Germany}

\author{Jonathan E. Kollmer}
\affil{University of Duisburg-Essen, Faculty of Physics, Lotharstr. 1-21, 47057 Duisburg, Germany}

\author{Tobias Steinpilz}
\affil{University of Duisburg-Essen, Faculty of Physics, Lotharstr. 1-21, 47057 Duisburg, Germany}

\author{Maximilian Kruss}
\affil{University of Duisburg-Essen, Faculty of Physics, Lotharstr. 1-21, 47057 Duisburg, Germany}

\author{Felix Jungmann}
\affil{University of Duisburg-Essen, Faculty of Physics, Lotharstr. 1-21, 47057 Duisburg, Germany}

\author{Tunahan Demirci}
\affil{University of Duisburg-Essen, Faculty of Physics, Lotharstr. 1-21, 47057 Duisburg, Germany}

\author{Jens Teiser}
\affil{University of Duisburg-Essen, Faculty of Physics, Lotharstr. 1-21, 47057 Duisburg, Germany}

\author{Gerhard Wurm}
\affil{University of Duisburg-Essen, Faculty of Physics, Lotharstr. 1-21, 47057 Duisburg, Germany}

\begin{abstract}

{In protoplanetary disks, zones of dense particle configuration  promote planet formation.}
{Solid particles in dense clouds} alter their motion through collective effects and back reaction to the gas.
The effect of particle-gas feedback with ambient solid-to-gas ratios $\epsilon > 1$ on the {stopping time of particles} is investigated.
In experiments on board the International Space Station we studied the evolution of a dense granular gas while interacting with air. We observed diffusion of clusters released at the onset of an experiment but also the formation of new dynamical clusters. The solid-to-gas mass ratio {outside the cluster} varied in the range of about $\epsilon_{\rm avg}  \sim 2.5 - 60$.
We find that the concept of gas drag in a viscous medium still holds, even if the medium is strongly dominated in mass by solids. However, {a collective factor} has to be used, depending on $\epsilon_{\rm avg} $, i.e. the drag force is reduced by a factor 18 at the highest mass ratios. Therefore, flocks of grains in protoplanetary disks move faster and collide faster than their constituents might suggest.
\end{abstract}


%


\section{Introduction}
In early phases of planet formation, dust-aggregates grow in collisions to millimeter size until growth stalls at the bouncing barrier \citep{BlumWurm2008, Zsom2010, Kelling2014, Johansen2014, Kruss2016, Demirci2017}. It was recently shown that one order of magnitude or more in size might be gained by charged aggregation \citep{Steinpilz2019b}. Promising ways to concentrate these cm to dm size pebbles further include particle-gas feedback, e.g. drag instabilities \citep{YoudinGoodman2005, JohansenYoudin2007, SquireHopkins18}. Particle over-densities might lead to gravitational collapse and the formation of planetesimals, eventually \citep{Johansen2007, Chiang2010, Simon2016}. With cohesion decreasing in importance, these particle systems become more and more granular. So small asteroids can be regarded as gravitationally bound rubble piles \citep{walsh2018}, and also comets might be set up as granular systems even though the volatile nature of water ice makes this case more complex \citep{Weissman2020}. 
During the stages of cometesimal and planetesimal formation through particle concentration, dense particle clouds are granular gases interacting with the molecular gas inside the protoplanetary disk.

Particle-gas interactions in protoplanetary disks are manifold. Depending on particle size and location, Epstein or Stokes drag laws apply \citep{Weidenschilling1977, Zsom2010, Johansen2014}. Due to their larger diameter, porous particles are more likely to be in Stokes drag regime \citep{Zsom2011, Okuzumi2012}.
This influences the radial drift of particles and determines the dust scale height, midplane sedimentation {and turbulent stirring} \citep{Weidenschilling1977,Dubrulle1995, Zsom2011, Birnstiel2016, Pignatale2017}.

In this work, we present results from experiments carried out under weightlessness on the International Space Station. The experiments show dynamical cluster formation and dissolution in a granular gas embedded in ambient air.
Under these conditions of long term microgravity, the motion of the granular gas is damped by gas drag.

{For a spherical particle in the Stokes regime, the friction force is}
\begin{equation}
\label{eq:StD}
F_\text{Stokes} = 6 \pi \eta_{mol} r v    
\end{equation}
where 
{$r$ is the radius of the particle, $\eta_{mol}$ is the molecular viscosity of the gas and}
$v$ is the \textit{relative} velocity between particle and the gas considered sufficiently far away from the particle.

{The gas-grain coupling time or friction time (of an individual test particle) is
\begin{equation}
\label{eq:tauf}
\tau_f = \frac{m v}{F_{\rm Stokes}} = \frac{2 r^2 \rho_{\rm particle}}{9 \eta_{mol}}.
\end{equation}
}
{The mass of a single particle is $m$ and the volumetric mass density of the particle is $\rho_{\rm particle}$.}

This is important for all relative particle motions, setting collision velocities and such. However, gas motion is not as easily accessible as particle motion and if particles move together in a clump, the focus is often on its absolute motion in a certain reference frame. 
For example, a  weakly permeable dense cluster of grains will sediment faster in the laboratory system than its individual components well separated from each other would do on their own. \citep{guazzelli2011, Schneider2019, Schneider2019b}.

{The gas-grain coupling time for a given dust grain increases within a denser environment or many particles close by \citet{Schneider2019b}. This corresponds to a decrease in the frictional drag force.}
The {collective behaviour} in our experiments {is expected} to depend on similar experiment parameters, e.g. particle to gas mass ratio {and volume filling factor. Former experiments{ of \citet{Schneider2019} and \citet{Schneider2019b} }relied on a rotational system to confine particles for long periods. The corresponding solid-to-gas ratios were below 1. In the microgravity experiments presented here, collective effects of particle gas interaction with solid-to-gas ratios of up to 60 were studied. }

The {collective effect of the particles on the gas} is calculated from the time-dependent kinetic energy for three different experimental settings.

\section{Experiments}

\begin{figure}[hbt]
    \centering
	\includegraphics[width=\columnwidth]{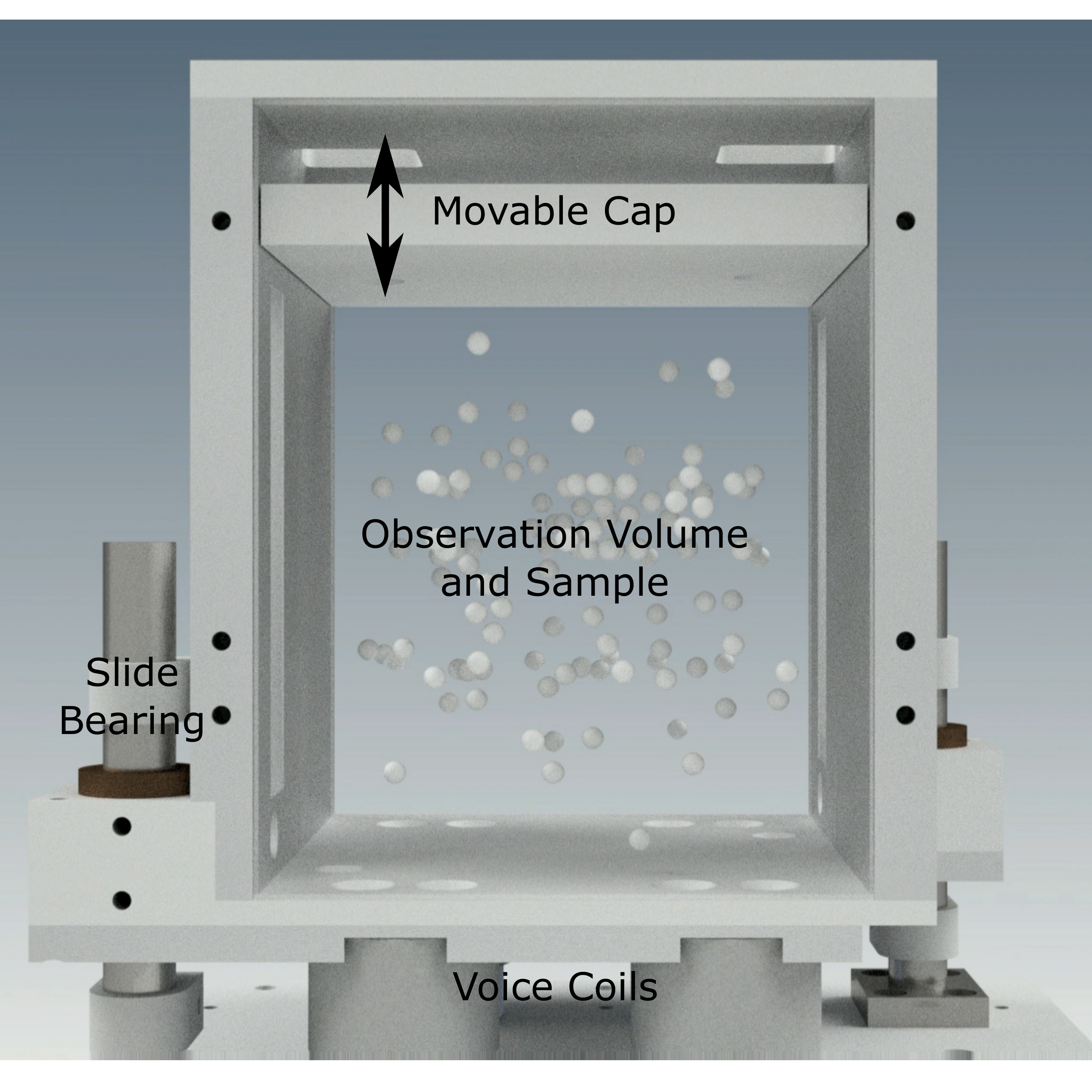}
	\caption{\label{Fig.exp}Sketch of the experiment chamber. Grains are entrained in gas at 1\,bar in a volume of maximum $5\times 5\times 5\, \mathrm{cm}^3$. The top wall can be moved down to 1 cm. Mesh covered slits on the left and right allow pressure adjustment on volume changes. The {entire} chamber can be agitated by voice coils and moves on horizontally mounted slide bearings. 
	}
\end{figure}

The experiments were carried out in the ARISE facility on the International Space Station in 2018 \citep{Steinpilz2019a}.
The essential part of this experiment consisted of a box filled with approximately 3500 spherical particles of $856 \pm 15 \, \rm  \mu m$ diameter.  The full technical and electrical design is described in \citet{Steinpilz2019a}. A sketch of the main part is shown in Fig. \ref{Fig.exp}.
The air pressure inside the chamber is $1 \, \rm bar$. The chamber can be shaken with typical frequencies in the range between 0.25 to $50\, \rm Hz$. The observed volume can be changed between $5\times 5\times 5\, \mathrm{cm}^3$ and $5\times 5\times 0.5\, \mathrm{cm}^3$ by moving a cap with up to a few mm/s. To balance the pressure inside the chamber, air can flow into the chamber or out of it through{ slits at} the left and right wall. These slits are covered with a $100 \, \rm \mu m$ mesh. The number of particles inside the chamber is always constant. As particles, we use SiO$_2$ (glass) spheres with a density of $2600\,\mathrm{kg}\,\mathrm{m}^{-3}$.
{For the largest volume, the average mass ratio of particles to gas is $\epsilon \approx 25$ and the volumetric filling factor is $\rm \Phi \approx 0.01 $}
The particles were imaged with up to 3280 by 2464 pixels with individual particles resolved by 25-30 pixels per diameter. Images are taken in bright field illumination and grains appear dark on a bright background as, e.g., seen in Fig. \ref{Fig.melt}.

\section{Data Analysis}

The main data were movies and snapshots of the particle clouds evolving.
Therefore, 2d images have to be analyzed. 
The images are 2d matrices with each pixel holding a gray scale. 
For well-separated grains, particle positions can be determined for different frames in a movie and velocities can be deduced. Knowing the mass of each grain, the kinetic energy can then be calculated for all grains.
However, in dense parts of the image, particles shade each other and cannot be tracked individually. Also, individual snapshots do not allow to trace velocities. 
It is therefore desirable to derive a quantity from a single image that is a function of the kinetic energy of all grains. It might not be intuitive that such a quantity exists. However, the number of clusters and individual grains is not independent of the dynamic state of the system. Therefore, the required information should be present in the gray scale values of individual images or its evolution with time. As one possibility, we introduce the Shannon values of 2d images here, a quantity of a granular system according to \citet{Shannon1948}. We note that this quantity is also known as informational entropy but we will refrain from using this term as it might imply thermodynamic associations in connection with granular or molecular gases, which are not intended.

\subsection{Shannon values of 2d images}

\begin{figure}[hbt!]
    \centering
	\includegraphics[width=0.99\columnwidth]{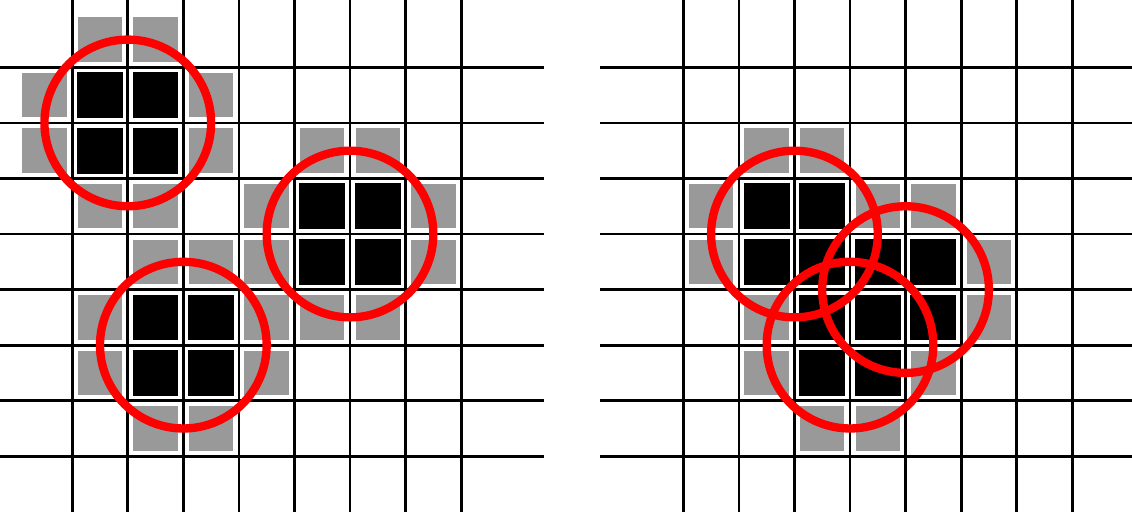}
	\caption{\label{Fig.SE}Illustration of the Shannon value of an image. Since the particles do not overlap in the left side image the larger amount of partially occupied border pixels leads to a more diverse distribution of gray scale values in the image and thus to a high Shannon value. In contrast, the overlapping disks on the right image form a cluster of black pixels with only a small amount of partially occupied border pixels and so the right image has a lower Shannon value.}
\end{figure}

In a gray scale image of absorbing particles, a cluster of grains forms a solid compact block. The image will show the cluster as a uniform dark area distinct from the uniform bright background. Some grains will be hidden beyond other grains though. If the particles in the image are all individuals, there is less overlap. This concept is illustrated in figure \ref{Fig.SE}.
To quantify this, the following value is calculated \citep{Shannon1948} 
\begin{equation}
H = -\sum_{i}p_i\log\left(p_i\right),
\end{equation}
where $p_i$ is the fraction of pixels with gray scale value $i$.
More specific, for an image with $X\times Y$ pixels and several $G$ possible gray levels, the Shannon value is calculated as
\begin{equation}
    H=-\sum_{i=1}^{G}\frac{N_{i}}{XY}\log\left(\frac{N_{i}}{XY}\right),
\end{equation}
where $N_{i}$ is the number of pixels with gray level $i$. 
{The possible grayscale values for each pixel range from 0 to 255 and each value is taken equivalently into account for the calculation of the Shannon value. In contrast to the schematic particles in Fig. \ref{Fig.SE} the particles used in the experiment are transparent. Therefore their inner part appears in light grayscales, while their rim appears to be darker. However, (partial) superposition of particles will lead to reduced grayscales, so the calculation of the Shannon function still holds.} 

\subsection{Shannon values and kinetic energy}

{Particles within a cluster do not move in the frame of reference of the cluster. Additionally, when the particles are agitated a larger part of the available volume is covered by particles, as they move {separately} in many directions. If the kinetic energy within the whole system is large (due to agitation), the particles are more dispersed and therefore will create a wider histogram of gray levels and a higher Shannon value. In case a large cluster forms, the cluster will remain rather stationary, so the total kinetic energy is rather low even in case of strong agitation. This is due to the fact that the agitation only reaches particles close to the moving parts of the chamber, while the clusters remain at rest.}

\begin{figure}[hbt!]
    \centering
	\includegraphics[width=0.99\columnwidth]{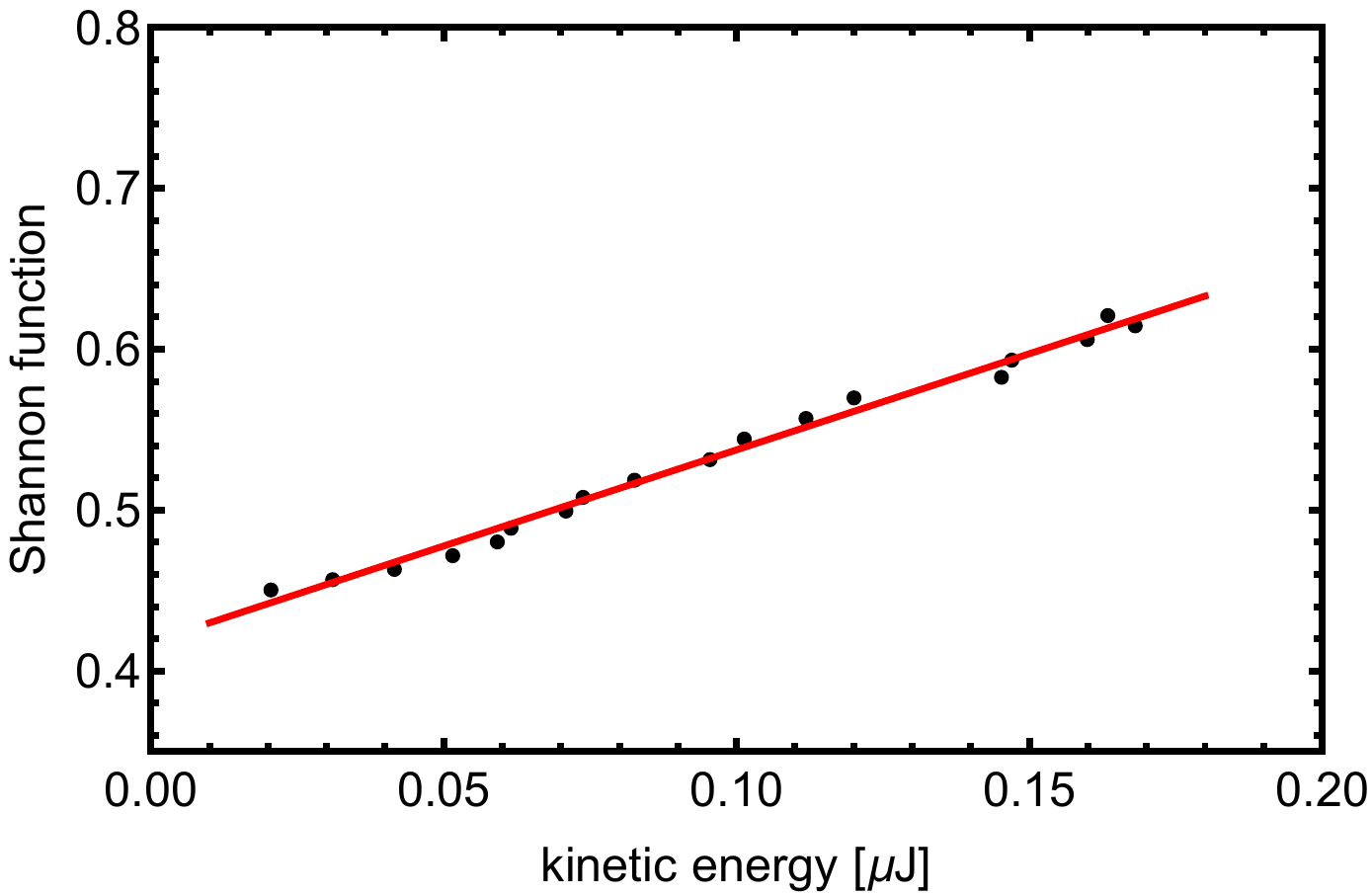}
	\caption{\label{Fig.s(E)}Normalized Shannon values depending on the mean kinetic energy per particles within the analyzed frame. The values were averaged {(simple moving average)} over 15 data points in time. The linear fit is plotted in red. {The fit parameters are $H_0 = 0.418 \pm 0.003$ and $\alpha_1 = 1.19 \pm 0.03 [\mu J^{-1}]$. $R^2$ is $0.999871$. }}
\end{figure}

As such, qualitatively, the Shannon function should be a measure for the kinetic energy {$E_\text{kin}$}. To test this hypothesis, we measured the kinetic energy of the particles by using Particle Tracking Velocimetry (PTV) for one experiment. We present data for the diffusing cluster case detailed below as an example. As can be seen in Fig. \ref{Fig.s(E)} the relation between the mean kinetic energy per particle and the Shannon values of the corresponding image is in very good agreement a linear dependence or
\begin{equation}
H = H_0 + \alpha_1 \cdot E_\text{kin}.
\label{HTrel}
\end{equation}
Here, $\alpha_1$ is a proportionality factor and $H_0$ a constant for the residual Shannon function. 
The particle velocities were estimated by evaluating PTV. Due to 2D
data, the kinetic energy is systematically too low by a factor 3/2, which is taken care of by the factor $\alpha_1$.

{The parameters $H_0$ and $\alpha_1$  vary for different experiments, depending on the initial particle distribution, chamber size and on agitation parameters of the experiment.} The big advantage of measuring the Shannon function instead of always using the PTV-method to deduce the kinetic energy is the lower time requirement in analyzing large data sets of images and the straightforward treatment of overlapping particles. We also tested part of the data of the two other cases discussed below and also found a linear relationship. We are therefore confident that the linear trend can be applied universally for the experiments. 

{While the Shannon value might be applicable to track cluster formation or cluster dissipation similar to cases of \citet{daisaka1999}, we only use it to determine the overall particle motion and estimate factors concerning the inter-particle collective behaviour.}
 
\subsection{{Collective drag} model}
\label{model}
The particle systems considered in our experiments are subject to different forces. The grains might be driven by collisions with an oscillating wall and drag by an external gas flow. We assume that these driving forces $F_\text{drive}$ are constant on first order.

Furthermore, all particles are assumed to be spherical with the radius $r$ and mass $m$. The damping of the grains is assumed to {occur} due to interaction between the particle and the surrounding air.

The Reynolds numbers $Re = \dfrac{\rho_{\rm air} \cdot r \cdot v}{\eta_{mol}}$ for all experiments are in the order of $10^{-2}$, {with the density of air $\rho_{\rm air} = \rm{1,204 \, kg\,m^{-3}}$}.

With the particle's velocity $v$ the drag force is given by Stokes' law {(Eq. \ref{eq:StD})}. The acceleration then is
\begin{equation}
\frac{dv}{dt}= \frac{F_\text{drive}}{m}-\frac{6\pi r \eta_{mol}}{m}v
\label{eq.moveit}
\end{equation}
Integration gives
\begin{equation}
v= \frac{F_\text{drive}}{6\pi r \eta_{mol}} + \left(v_0-\frac{F_\text{drive}}{6\pi r \eta_{mol}}\right)\exp\left(-\frac{6\pi r \eta_{mol}}{m}t\right),
\end{equation}
{with initial velocity $v_0$.}
The kinetic
energy of $N$ particles is
\begin{equation}
\begin{aligned}[c]
E_\text{kin}=\frac{Nm}{2}v^2.
\end{aligned}
\end{equation}
We now switch to the Shannon function, which is then also essentially an exponential function in time and can be expressed as
\begin{equation}
\begin{aligned}[c]
H &= H_0+\alpha_1\frac{Nm{F_\text{drive}}^2}{72\pi^2r^2 \eta_{mol}^2}\\
&+\alpha_1\frac{NmF_\text{drive}}{6\pi r \eta_{mol}}\left( v_0-\frac{F_\text{drive}}{6\pi r \eta_{mol}}\right)\exp\left(-\frac{6\pi r \eta_{mol}}{m}t\right)\\
&+\alpha_1\frac{Nm}{2}\left( v_0-\frac{F_\text{drive}}{6\pi r \eta_{mol}}\right)^2\exp\left(-\frac{12\pi r \eta_{mol}}{m}t\right).\\
\end{aligned}
\label{Htexpand}
\end{equation}

{This result is consistent with \citet{Masiuk2008}, who also assumed the increase of the Shannon function to be exponential in time when using the Shannon function to determine the mixing entropy of a driven granulate from image data.}
Abbreviating the constants to $a$, $b_1$, $b_2$, and $c$ this yields
\begin{equation}
    H(t) =  a+b_1\exp(-c t)+b_2\exp(-2 c t).
    \label{Ht}
\end{equation}
{Parameters $b_1$ and $b_2$ are positive, when particles are diffusing, or negative, when particles are clustering.}
Important here is that the fit value $c = \frac{6\pi r \eta_{mol}}{m}$ depends on the {well-known parameters $r$, $m$ and $\eta_{mol}$}. This fit value is independent of the underestimation of the velocity {and experimental parameters like $H_0$ or $\alpha_1$.}

{The Stokes model considers only one single sphere. Considering dense particle clouds and back-reaction on the gas flow, this is not valid anymore \citep{guazzelli2011, Schneider2019, Schneider2019b}. The calculated base value for $c$ is $0.17 s^{-1}$, which is significantly smaller than the measured values (see section \ref{sec:ExpSet}). Therefore, we introduce a dimensionless collective correction factor $\chi$.}
{Using Eq. \ref{eq:tauf}, the fit parameter c in our model is now}
\begin{equation}
    c  = \dfrac{1}{\chi \cdot \tau_f},
    \label{eq:fitparaC}
\end{equation}
{with $\chi = 1$ for single particles and $\chi \neq 1$ relating to collective effects.}

This is our model prediction for the time-dependent Shannon function in driven particle systems with gas friction and the way we determine {the collective factor $\chi$}. We analyzed data from three specific experiments to correlate the collective factor to the average solid-to-gas mass ratios $\epsilon$. 
{We do not measure the gas velocity and can not determine the relative velocities between particles and fluid. Our estimations {on $\chi$ are} based on the acceleration of particles in the laboratory system. It's more than likely, that particles back-react on the gas and drag the gas with themselves. This would also increase the measured {collective factor} and matches the picture of mechanisms in drag instabilities \citep{Jacquet2011, SquireHopkins18}.}

{We expect the feedback of the gas to scale with the solid-to-gas ratio to ensure momentum conservation \citep{YoudinJohansen2007}.}
{The solid-to-gas ratio is estimated by counting all particles $N$ not located inside the cluster averaged over the time of the respective experiment. It is proportional to the volume filling factor $\Phi$. }
\begin{equation}
    \label{phi}
    \Phi = \dfrac{N\cdot V_{\rm particle}}{V_{\rm chamber}}
\end{equation}

\begin{equation}
    \label{epsilon}
    \epsilon= \Phi \cdot \dfrac{\rho_{\rm particle}}{\rho_{\rm air}}
\end{equation}

{$V$ is the volume of the particle or the chamber, $\rho$ is the volumetric mass density of the particles or the air.}

\vspace{1cm}

\section{Experimental settings}
\label{sec:ExpSet}

In the first experiment, a cluster, initially at rest, is dispersed by the airflow in the system as the cap is moved. The second experiment also has a cluster initially at rest but energy is injected by a short pulsed motion of the experiment chamber walls. The third type of experiment is the formation of a dynamic (non-sticking but locally dense) particle cluster in a continuously vibrated case.

\subsection{Cluster diffusion}

\begin{figure*}[]
    \centering
	\includegraphics[width=2\columnwidth]{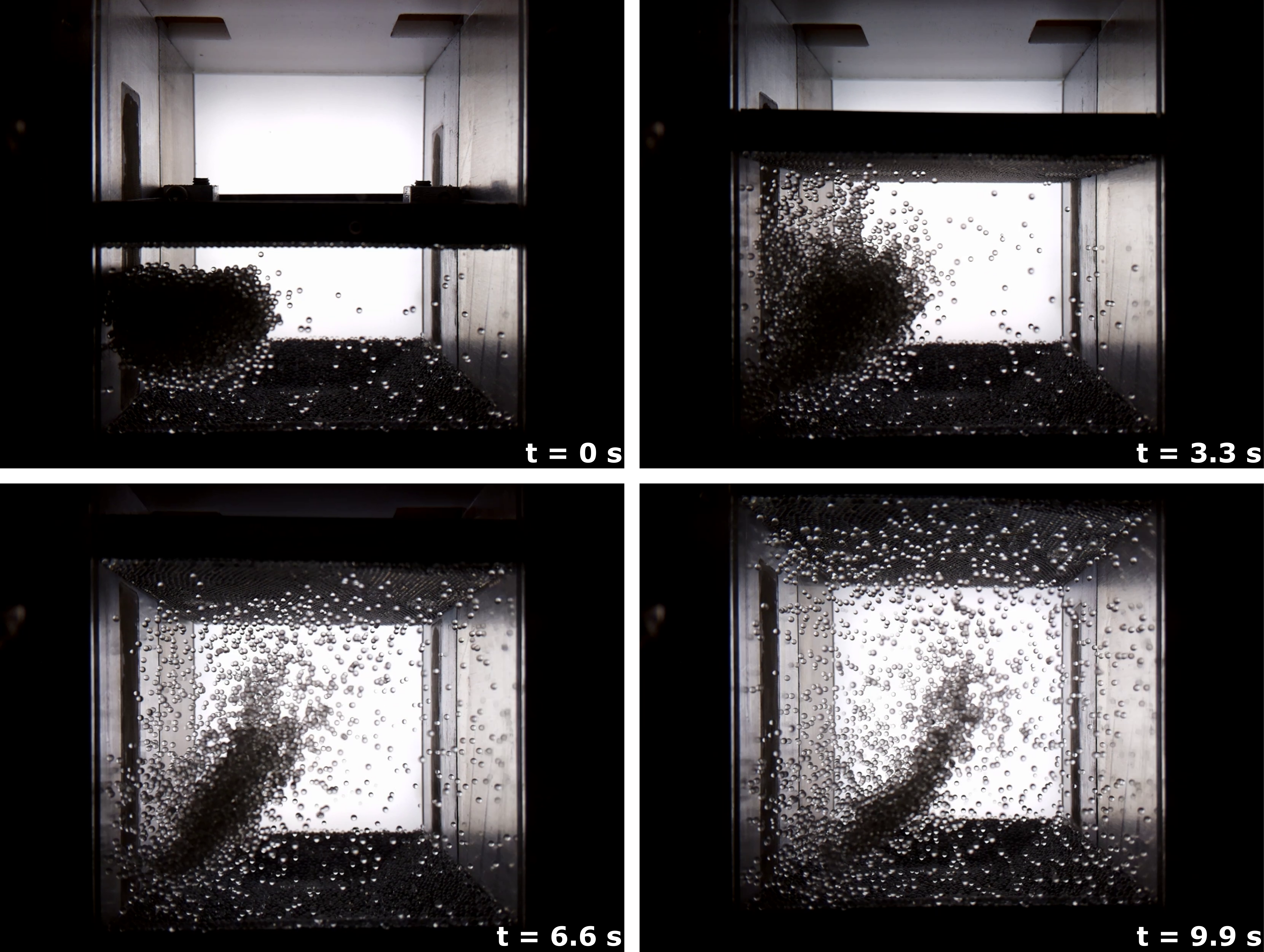}
	\caption{\label{Fig.melt}Sequence from a giant cluster diffusion. The cap inside the observation chamber is moved upwards creating a gas flow. This gas drag forces the cluster to melt. The observation chamber has a dimension of 5 cm on each side.}
\end{figure*}

\begin{figure}[]
    \centering
	\includegraphics[width=0.99\columnwidth]{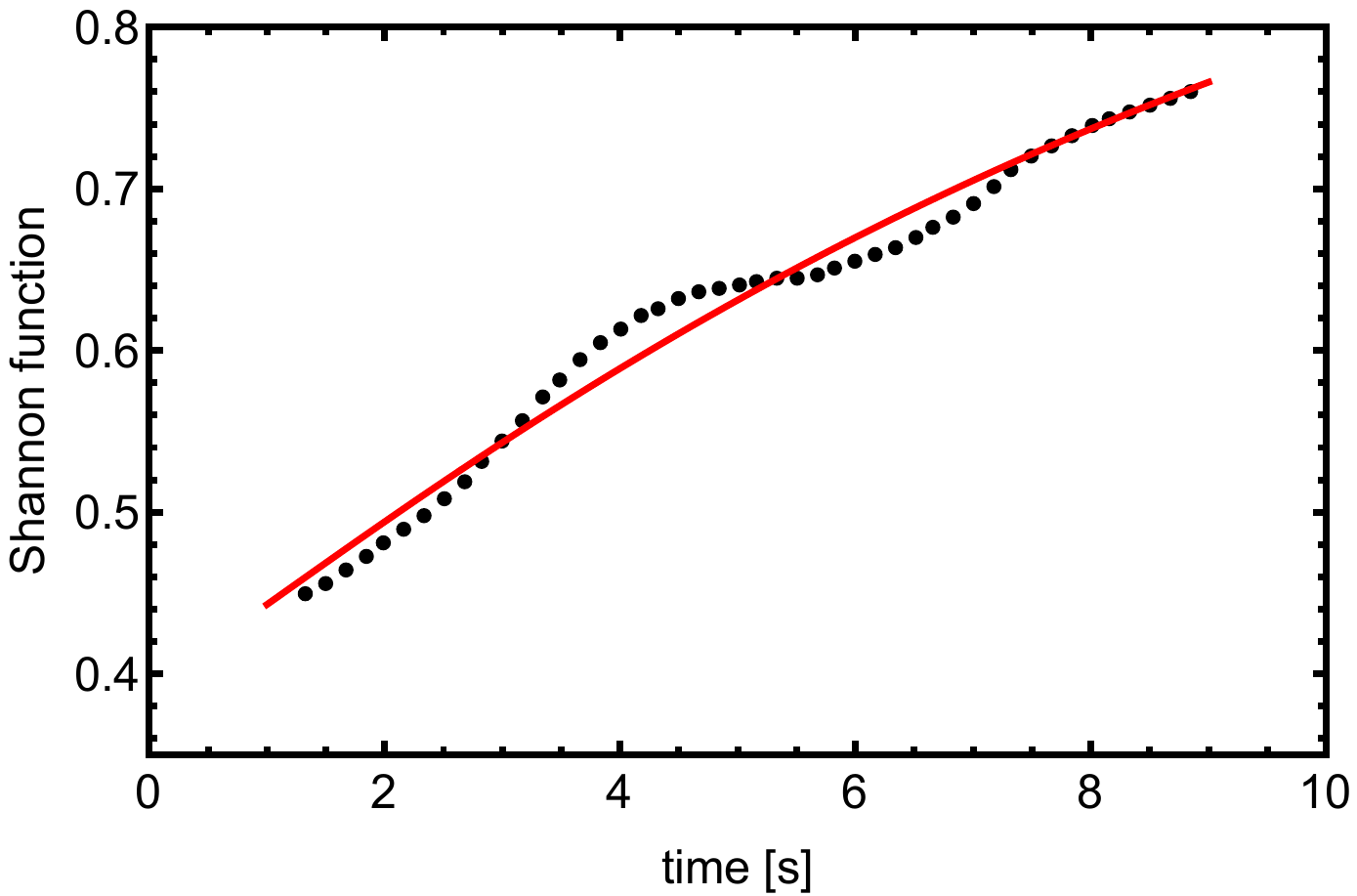}
	\caption{\label{Fig.entropy}Normalized Shannon function over time for the diffusing cluster in Fig. \ref{Fig.melt}. The line shows an exponential fit to the data. The Shannon values were calculated for a constant area in the analyzed image without the movable cap and normalized by the constant value of the  exponential. The values were averaged {(simple moving average)} over 15 data points in time.}
\end{figure}

The first and simplest experiment for cluster diffusion is carried out with 50 Hz vibration while the cap is moved.
{The cluster forms during a pre-experiment routine by residual gravity. The forming cluster is bound together by contact forces or electrostatic forces. By increasing the container volume from $62.5 \, \rm{cm^3}$ to $125 \, \rm{cm^3}$ in $10 \, \rm s$} we create an airflow to which the particles couple and by which they move around.  {The }initially very dense cluster of grains is diluted and the particles form a suspension in the surrounding air. The cluster is surrounded by almost no particles initially but the number density increases over time. Sample images from the video taken during one of these experiments are shown in Fig. \ref{Fig.melt}.
Fig. \ref{Fig.entropy} shows the Shannon values as a function of time for this case and an exponential fit as we would expect from the model. 
Resulting from the fit,{ a collective factor of $\chi = 1.8$} can be calculated. 

Small and sometimes periodic deviations in the Shannon function occur. These might be caused by particles that leave or enter the observation area of the inner 50 \% of the image used. Also, the particle number density varies with time which would introduce a systematic deviation in detail, which we neglect here.

\subsection{Cluster diffusion with agitation}

\begin{figure*}[]
    \centering
	\includegraphics[width=2\columnwidth]{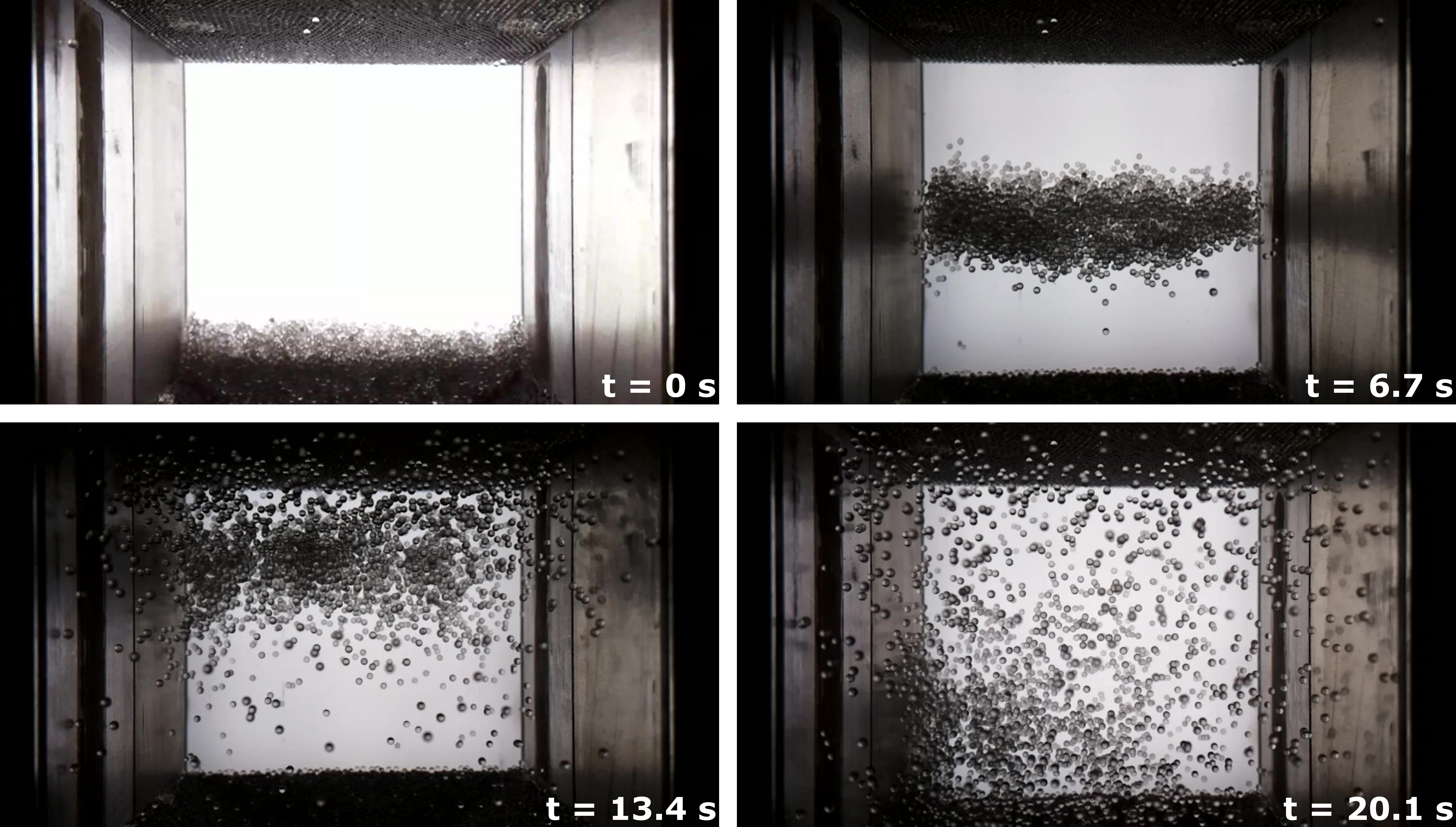}
	\caption{A cluster diffuses as it bounces forth and back between walls (top and bottom wall) which are moved with $1\, \rm Hz$ (step-function). The time between the images is $6.7\, \rm s$.}
	\label{Fig.shake}
\end{figure*}

\begin{figure}[]
    \centering
	\includegraphics[width=0.99\columnwidth]{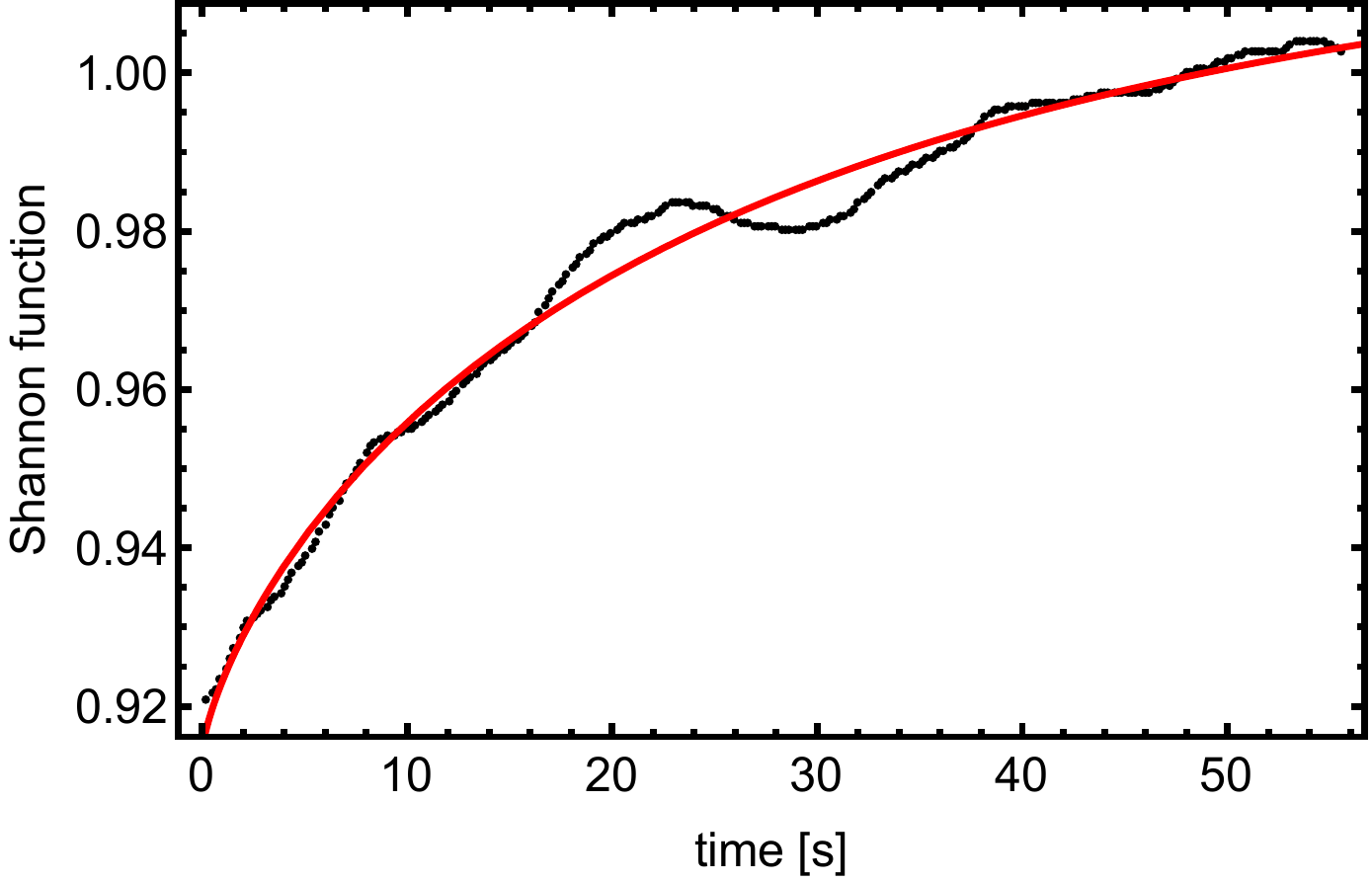}
	\caption{\label{Fig.entropyshake}Time-dependent Shannon value for cluster diffusion with agitation ({simple }moving average of 100 data points in time).}
\end{figure}

{The second experiment starts with a cluster formed at the bottom of the observation volume by residual gravity}. During the diffusion experiment, we used an agitation by moving the chamber stepwise with a frequency of $1\, \rm Hz$ and an amplitude of a few millimeters. Sample images are shown in Fig. \ref{Fig.shake}.
Again, the initial cluster can be observed to dissolve. In contrast to the first experiment, the cluster is not as dense and also produces more individual particles.
However, the general behavior of the Shannon value is still exponential in time when applying a moving average (see Fig. \ref{Fig.entropyshake}). Again, the exponential fit as expected by the model is shown. 
The {collective factor} determined in this case is $\chi = 3$.
This is larger than in the first case while at the same time the average solid-to-gas ratio is higher.


\subsection{Dynamic clustering due to agitation}

\begin{figure*}[]
    \centering
	\includegraphics[width=2\columnwidth]{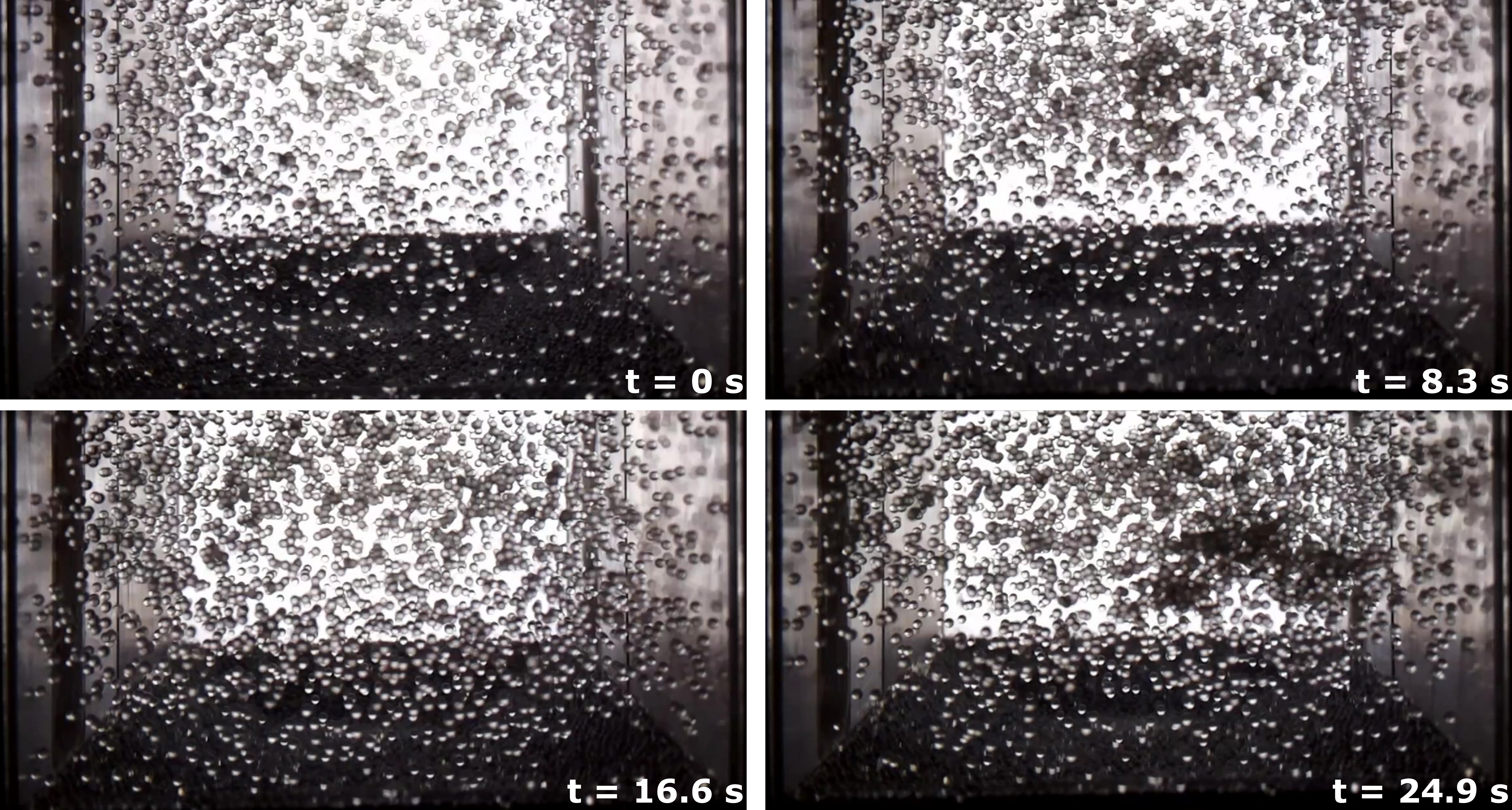}
	\caption{\label{Fig.clustering} A dynamic cluster forms in the observation chamber. }
\end{figure*}

\begin{figure}[]
    \centering
	\includegraphics[width=\columnwidth]{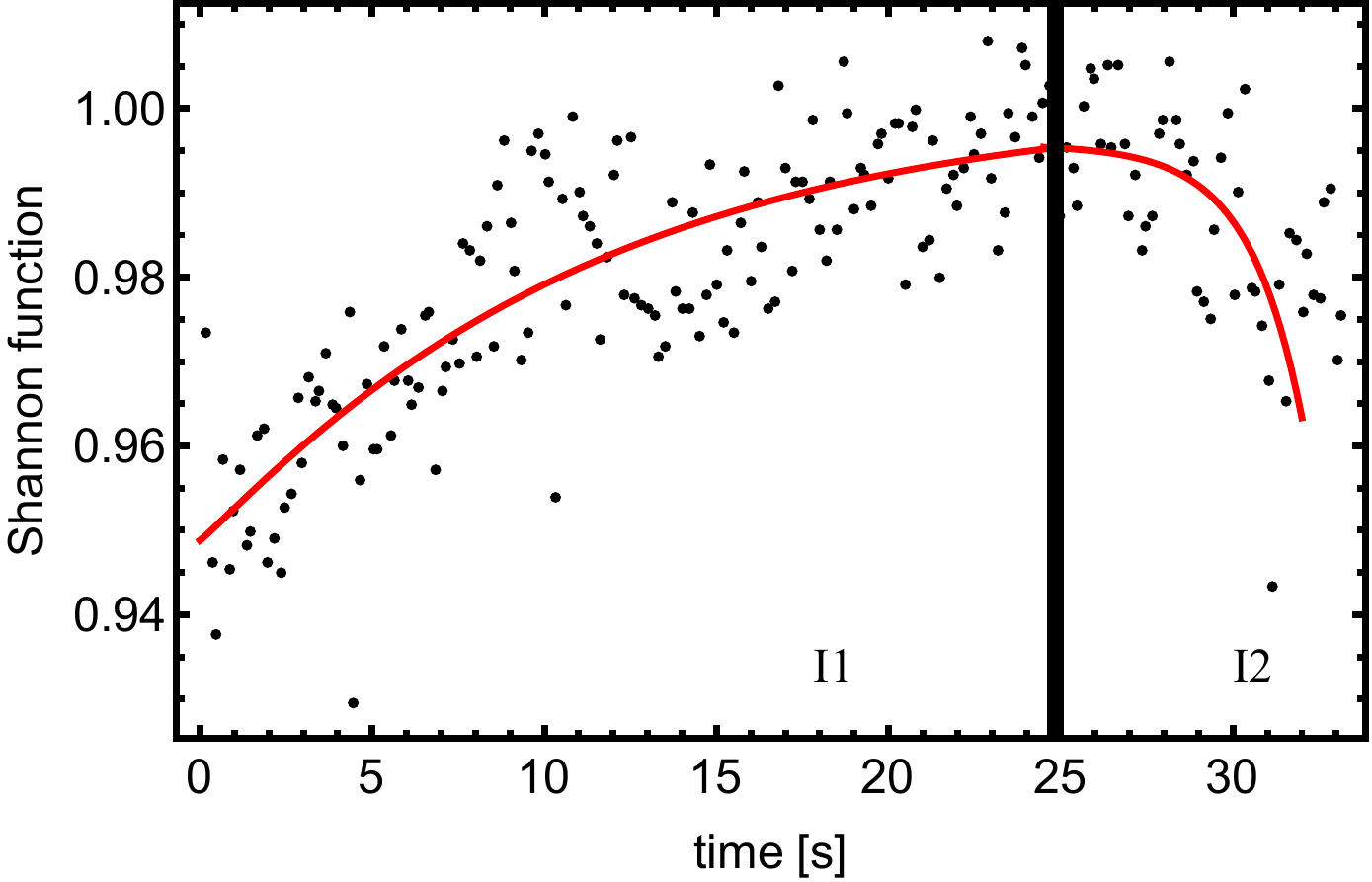}
	\caption{\label{Fig.entropyclustering}Time-dependent Shannon values for clustering with agitation. The first interval (I1) represents the agitated system without cluster formation. Within the second interval (I2) a cluster forms for the agitation parameters.}
\end{figure}

The third and last experiment described here shows clustering due to agitation.
A sample image sequence for this experiment is shown in Fig. \ref{Fig.clustering}.
To achieve the conditions for clustering the system is shaken with a frequency of about $50\, \rm Hz$ while the cap is lowered continuously to a volume of $5\times 5\times 2\, \mathrm{cm}^3$. Only then, a denser part within the system forms. The formation of a dynamic cluster is an interesting feature. However, it is well known that clusters only form for specific agitation{and particle collision parameters, e.g. the size of the system compared to the diameter of the particles, the volume filling factor of the system and the amplitude and frequency of the wall oscillations \citep{Meerson2004,Opsomer2012,Opsomer2014,Noirhomme2015}}. The experiment might add a new data point concerning this topic but this is not the focus of this paper.
Here, it is important that with this heavy driving, the particles are essentially distributed homogeneously inside the chamber and the solid-to-gas ratio is close to maximum.
The corresponding Shannon function is shown in Fig. \ref{Fig.entropyclustering}.  
Pinning down the moment of cluster formation is not straight forward by
viewing the images due to the large number of particles and local clusters that form statistically at all times.  However, the Shannon function constrains the time of cluster formation very well. 

As the system is agitated, the Shannon values increase (I1). When a cluster begins to form, the Shannon function decreases again (I2). The time dependence of both - the increasing and decreasing part - can be described again by exponentials. 
The {collective factor for this experiment is $\chi = 18$} for both parts of dynamic clustering, which is {the largest of the three experiments} and again related to a further increase in the average solid-to-gas ratio.

\section{Collective motion and back reaction}
\label{Sec:CollMo}

{The determined collective factors deviate from the base value of $\chi = 1$ valid for single particles up to an order of magnitude.
The dilute system of a diffusing cluster is closest to the isolated particle case while the value for the dense cloud is strongly increased.}

{Particle-particle collisions might act as a damping mechanism. }
The collision timescale between two grains is 
\begin{equation}
\tau_c = \frac{1}{\sigma_c n v}.
\end{equation}
Here, $\sigma_c = 4 \pi r^2$ is the collisional cross-section, $n$ is the particle number density and $v$ the relative velocity of the grains. Taking $v =$ 2 mm/s, 1.7 mm/s, and 2.5 mm/s as collision velocities
{ determined by PTV, $n=$ $3.4\cdot 10^{6} {\rm m^{-3}}$, $13\cdot 10^{6}{\rm m^{-3}}$, and $81\cdot 10^{6} {\rm m^{-3}}$ as particle number density,}
the collision times in all three experiments are $\tau_c =$ 63 s, 20 s  and 2 s, respectively. 

{Using Eq. \ref{eq:tauf},} the friction times are then $\tau_f = 6\, s$ for the three experiments.   
The collision time is larger than the friction time in the first two cases and therefore does not influence the damping. {Collisions with the walls are accounted for by the driving term of eq. \ref{eq.moveit}. Collision times} {become comparable to the friction time for the third experiment though and collisional damping has to be considered in more detail. 
\citet{Steinpilz2019a} measured a coefficient of restitution of about $c_r \sim 0.5$ in the given experiments. Therefore, 75\% of the energy of a particle is lost after a collision. On an (uncorrected) friction time a fraction of $(1/e)^2$ or 86 \% of the kinetic energy is dissipated. If we consider also that collisions slow down the particles and collision times for a next collision increase and as gas drag already slows down particles in between collisions, there is about one collision per (uncorrected) friction time. Therefore, as worst case, collisional damping might lead to an overestimate of the gas drag by a factor of approx. 2. However, we also have to keep in mind that already in the more dilute cases, back reaction is very efficient and might effectively increase the friction time. Keeping this in mind, gas drag very likely also dominates in the third experiment. In any case, the general trend is valid.}

In summary, we get the following results:

\begin{table}
    \centering
    \begin{tabular}{r|r|r}
            Experiment & $\epsilon_{\rm avg}$  & $\chi$ \\
            \hline
            \hline
            
            cluster {diffusion}  &   2.5  &    1.8 \\
            \hline
            cluster {diffusion}  &   10  &   6 \\
            with agitation      &  {}    &   {}    \\
            \hline
            dynamic clustering  &   60 &   18  

    \end{tabular}
    \caption{Experimental parameters and collective factors}
    \label{tab:conclusion}
\end{table}

\begin{figure}[]
    \centering
	\includegraphics[width=\columnwidth]{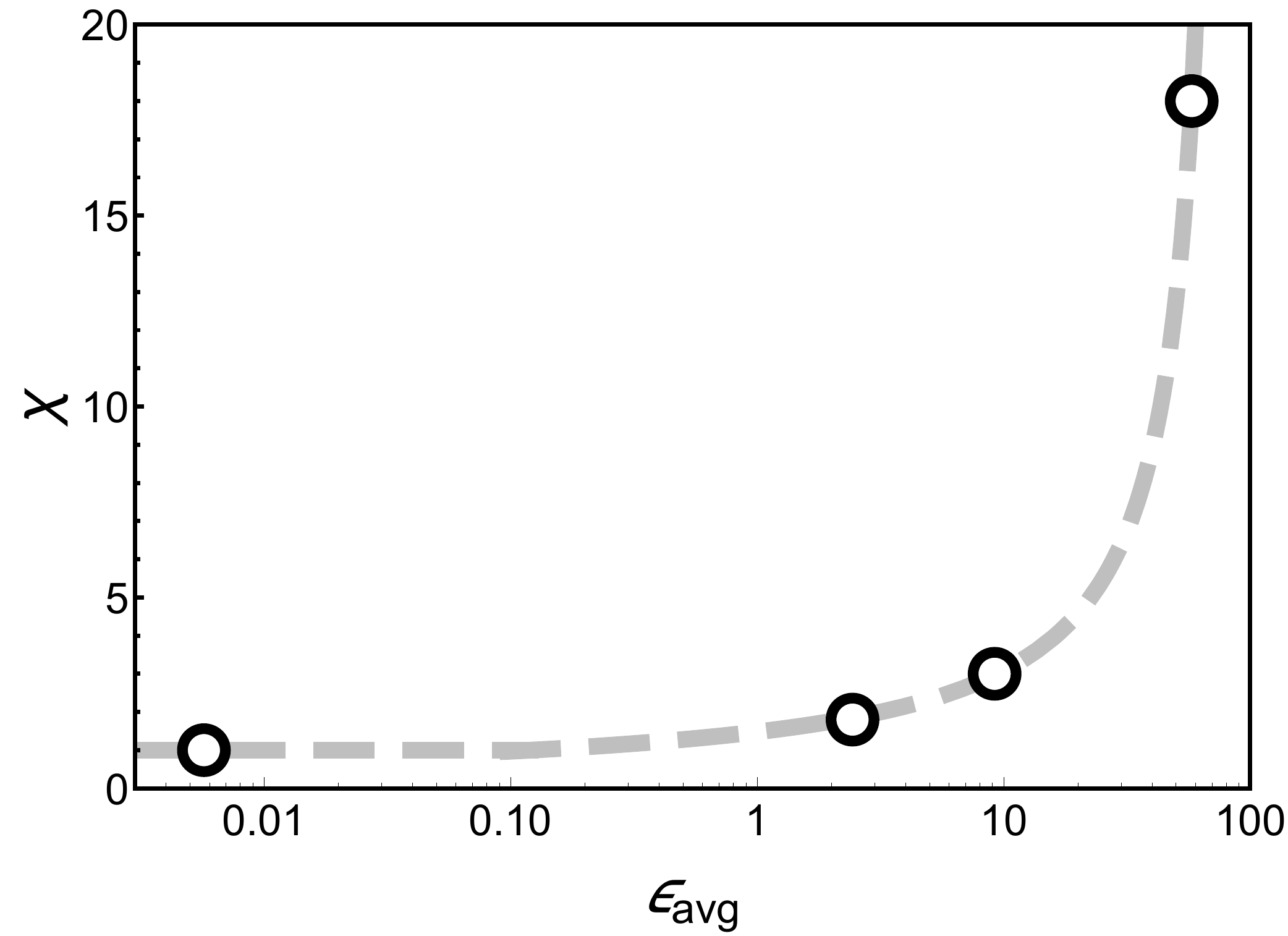}
	\caption{Correction factor in dependence of the average solid-to-gas ratio. The leftmost point corresponds to the estimation of a single sphere in our experimental chamber with standard drag laws ($\chi = 1$). The fit function is $\chi(\epsilon_{\rm avg}) = (0.68 ^{\pm 0.01} - 0.153^{\pm 0.003} \cdot  \log(\epsilon_{\rm avg}))^{-1}$}
	\label{fig:chieps}
\end{figure}

Overall, our findings imply that the {standard drag force model considering only single particles} is largely overestimated. {The stopping time is increasing with increasing ambient $\epsilon_{\rm avg}$ (Table \ref{tab:conclusion} and Fig. \ref{fig:chieps})}. This is consistent with  \citet{Schneider2019} and \citet{Schneider2019b} who find that collective behavior of particle clouds is induced in high density clouds. 

The back reactions from solid particles to the gas were also increasing with solid-to-gas ratio in \citet{Schneider2019b}. 
In their case, it was only triggered at a certain, minimum {solid-to-gas ratio depending on the Stokes number (particle stopping time multiplied by rotation frequency of the system)}. 
{Due to different experimental setups, a one to one comparison for the minimum solid-to-gas ratio cannot be given for the experiments presented in this work.
Extrapolating the trend in Fig. \ref{fig:chieps}, the critical solid-to-gas ratio is $0.14$.
The collective factors were always {larger} than 1, showing that $\epsilon_{\rm experiment} > \epsilon_{\rm crit}$ and therefore, collective behaviour is always present.}

It remains debatable, that $\chi$ in our model might not influence the stopping time, but only one of the parameters in Eq. \ref{eq:tauf}, namely $\eta_{mol}$, $r$ and $m$ or $\rho_{\rm particle}$.
In fact, there are models of effective viscosity and effective aerodynamic radius.

Generally, the viscosity of suspensions of solid particles is expected to increase with the volume filling factor, e.g. the Einstein viscosity and further models \citep{brinkman1949, Einstein1956, Hsueh2005, Petford2009, Mueller2010}. In contrast to these findings, our experiments with volume filling factors $\Phi < 10^{-1}$ show an increased stopping time and therefore, a decreased effective viscosity for denser particle configurations as detailed below.
This is not a contradiction but hinges on the definition of viscosity.

 According to \citet{Einstein1956}, the viscosity increases but if Eq. \ref{eq:StD} was used as a definition in the laboratory system, the effective viscosity would be lower. This is due to the fact that the aggregate drags the gas along and the relative velocity between gas and particle is smaller than the absolute sedimentation velocity. This is not considered in the laboratory reference frame. Still, the definition of viscosity, which might be called effective viscosity then, is a useful concept. It allows the treatment of particle motion without the need to figure out the details of the back reactions to the gas.

An effective radius as, e.g used in the aerodynamic diameter does not take into account any collective effects, but is commonly used in aerosol physics to describe and compare particles of different shape, porosity or volumetric mass density \citep{decarlo2004}. The aerodynamic diameter can be larger or smaller than the physical diameter. Relating the collective effect only on the particle size might sufficiently describe that small particles behave like larger particles. This might also shift the transition from Stokes to Epstein regime according to the change in effective size.

Our statements, in this case, are different from other constraints as follows:

\begin{itemize}
\item The {stopping time of} particles in a dense cloud can still be described by {common drag laws taking into account the collective factor $\chi$} even if the solids dominate in mass by a factor of 60.
\item This refers to a motion of grains which is widely random in direction, i.e. does not have a systematic rotation or drift.
\item With an {ambient} solid-to-gas ratio of $\epsilon_{\rm avg}=2.5$ the {stopping time is increased by a factor of  about 2}.
\item At very high solid-to-gas ratios of several tens, gas drag is still a viable concept but is reduced by a factor of 18. At these high solid-to-gas ratios {the Streaming Instability might cause particle} clumping {in protoplanetary disks, that is potentially followed by} gravitational collapse.
\item With a more than linear decrease with $\epsilon_{\rm avg}$ the drag in clouds up to $\epsilon_{\rm avg}=0.1$ might well be described by the regular {gas drag laws}. 

\vspace{.5cm}
\end{itemize}

\subsection{Implications on Planet Formation}

\begin{figure}[]
    \centering
   
	\includegraphics[width=\columnwidth]{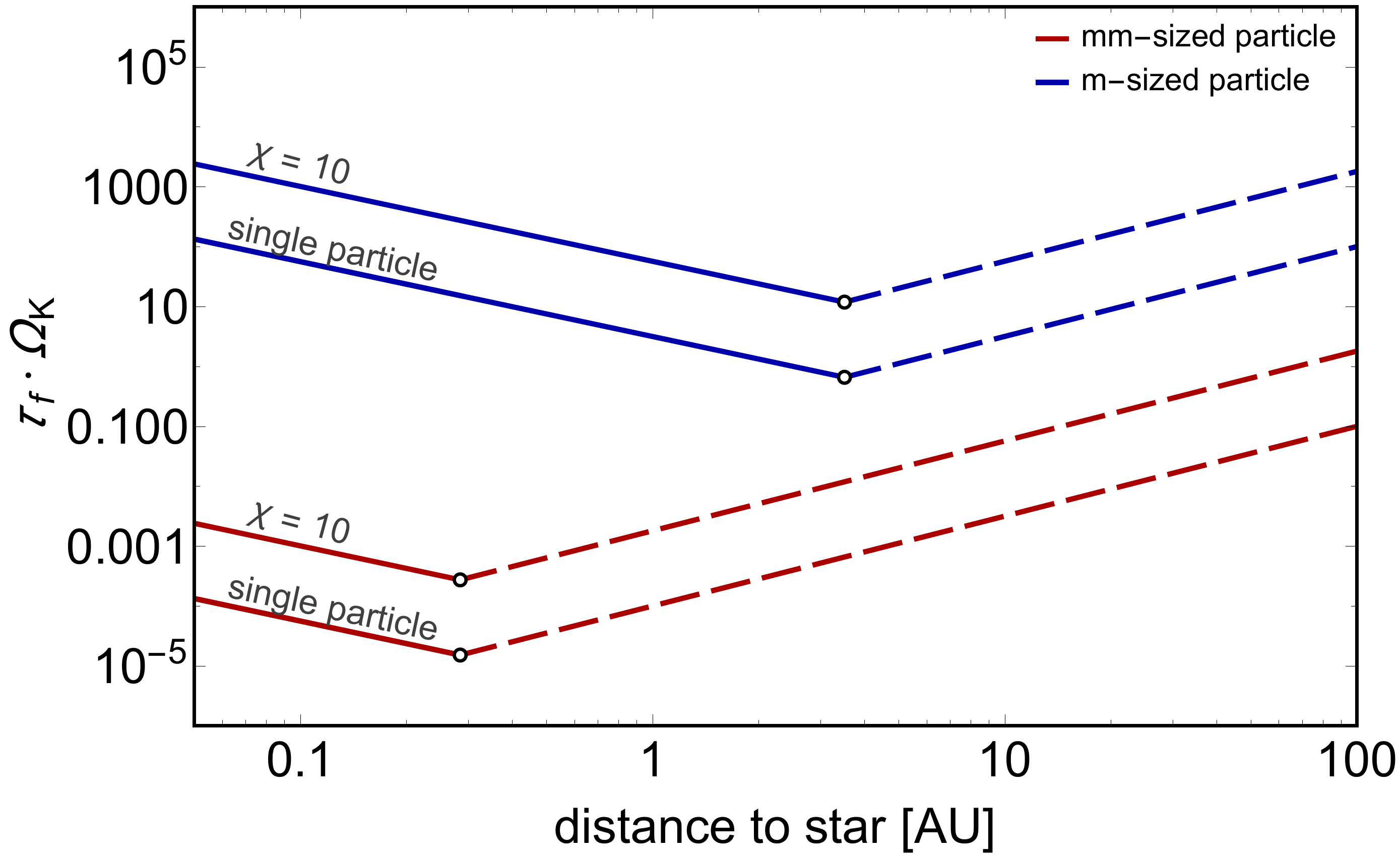}
	
	\caption{Shift of the dimensionless stopping time $\tau_f \cdot \Omega_{\rm K}$ for mm-sized and m-sized particles. The solid line marks the Stokes drag regime, the dashed line shows the drag in Epstein regime. Following the interpolation of Fig. \ref{fig:chieps}, $\chi = 10$ corresponds to $\epsilon = 42$.}
	 \label{fig:tausemi}
\end{figure}

The implications of a change in stopping time for particles in close configurations are manifold. Figure \ref{fig:tausemi} shows the calculated stopping time of mm- and m-sized particles using the standard stopping time and in a configuration with $\chi = 10$ ($\epsilon=42$). Although our findings might not hold one to one in Epstein regime, collective effects should influence the transition regime for Knudsen number $Kn < 10$.
The stopping time of particles in close configurations is generally larger than the stopping time of single particles. Effectively, e.g. mm-sized particles behave like cm-sized particles.

How drag instabilities as in \citet{YoudinGoodman2005, Auffinger2017, SquireHopkins18} or \citet{gonzalez2017} are influenced by the increased stopping time and the collective correction factor $\chi$ can only be discussed qualitatively and requires further investigations. 
For models of planet formation relying on drag and particle-gas interaction like the Streaming Instability, the effect of increasing stopping time in dense particle configurations might be of great importance. Small particles in a dense configuration increase their stopping time and therefore, the minimum solid abundance needed to start the Streaming Instability is potentially decreased \citep{Carrera2015, Yang2017}. Furthermore, the concentration of small particles with larger stopping time is faster \citep{Yang2017}.

\citet{Zsom2011} and \citet{Okuzumi2012} state that porous particles might break the radial drift barrier in Stokes regime. Our findings might further increase this effect. Porous particles are more likely to form in dense particle configurations via hit and stick.
Increasing the physical size via porous aggregation and simultaneously increasing the aerodynamic size of the particle by collective effects will most likely assist in overcoming the radial drift barrier.

In global simulations of protoplanetary disks, an adaption in the definition of viscosity is also used. 
In $\alpha$-disk models a viscosity due to turbulence is introduced \citep{shakura1973, frank2002}. Any interpretation of our finding as a change in effective viscosity will only change local simulations where particles are influenced by Stokes drag.

\vspace{1cm}

\section{Conclusion}

{We studied the evolution of granular systems where particles dominate in mass over gas by a factor of up to about hundred. With the use of a collective correction factor $\chi$}, the concept of gas drag based on {single particles} is still viable. {The stopping time of particles} is {increased}, in our parameter range by a factor up to 18. 

Our results here are a minimum correction for random motions, e.g. for particle evolution driven by local turbulence  or diffusive motion.
While solid-to-gas ratios start with a canonical value of 1/100 in protoplanetary disks, sedimentation and other processes rapidly increase this ratio to beyond one. 
At the highest densities, Stokes drag is significantly reduced. These results show that {particles in swarms have larger stopping times than single particles and therefore, are less coupled to the gas and} move faster than expected.

Collective motion of solids and back reaction to the gas in dense particle clouds are complex. However, the use of a collective correction factor in drag models of dense particles clouds with high solid-to-gas ratios should be taken into account.

\section{Acknowledgments}
This work is supported by the German Space Administration (DLR) with funds provided by the Federal Ministry for Economic Affairs and Energy (BMWi). Access to the International Space Station was provided under grant 50 JR 1703. Preparatory work was supported by grant 50 WM 1542. F. Jungmann and T. Demirci are supported by grants 50 WM 1762 and 50 WM 1760, respectively. M. Kruss is funded by DFG grant WU 321 / 14-1. N. Schneider is funded by DFG grant WU 321 / 16-1.
We appreciate the constructive review by the two anonymous referees.

\bibliography{references.bib}

\end{document}